\documentclass[final,3p,times,twocolumn]{elsarticle}

\usepackage{amssymb}

\journal{Journal of Magnetism and Magnetic Materials}

\begin{document}

\begin{frontmatter}



\title{
\begin{minipage}[t]{7.0in}
\scriptsize
\begin{quote}
\leftline{{\it J. Magn. Magn. Mater.}, in press.
}
\end{quote}
\end{minipage}
\medskip
Stoner Magnetism in an Inversion Layer}


\author{D. I. Golosov}
\ead{Denis.Golosov@biu.ac.il}
\address{Department of Physics and the Resnick Institute, Bar-Ilan 
University, Ramat-Gan 52900, Israel.}

\begin{abstract}
Motivated by recent experimental work on magnetic properties of Si-MOSFETs, 
we report a calculation of magnetisation and susceptibility of electrons in 
an inversion layer, taking into account the co-ordinate dependence of electron 
wave function in the direction perpendicular to the plane. It is assumed that 
the inversion-layer carriers interact via a contact repulsive potential, 
which is treated at a mean-field level, resulting in a self-consistent change 
of profile of the wave functions. We find that the results differ 
significantly from those obtained in the pure 2DEG case (where no provision 
is made for a quantum motion in the transverse direction). Specifically, the 
critical value of interaction needed to attain the ferromagnetic (Stoner) 
instability is decreased and the Stoner criterion is therefore
relaxed. This leads to an increased susceptibility and ultimately to a 
ferromagnetic transition deep in the high-density metallic regime. 
In the opposite limit of low carrier densities, a phenomenological treatment 
of the 
in-plane correlation effects suggests a ferromagnetic instability
above the metal-insulator transition.  
Results are discussed in the context of the available experimental data.
\end{abstract}


\end{frontmatter}


\section{Introduction}
\label{sec:intro}

Recently, there has been much interest in the unusual magnetic and 
magnetotransport properties of 2-dimensional (2D) electron gas (2DEG) at 
low densities, as exemplified by the inversion layers in Silicon 
metal-oxide-semiconductor field effect transistors (Si-MOSFETs)
\cite{Ando82, Kravchenko04}, 
or by GaAs quantum wells. Specifically, indirect magnetisation and 
susceptibility measurements \cite{Reznikov03,Kravchenko06} indicate an 
increase of low-field magnetic susceptibility with lowering the 2-dimensional 
carrier density $n$ toward the critical value $n_c$, corresponding to the 
metal-insulator transition (MIT). This might imply the presence of a 
ferromagnetic transition at or near MIT \cite{Kravchenko06}, although
the situation remains uncertain due to experimental difficulties. 

On the theory side, the possibility of ferromagnetism in a 2DEG at low density 
was raised in the numerical investigations\cite{Tanatar89,Attaccalite02}. 
Later, a divergence in susceptibility with decreasing $n$ toward $n_c$ 
was reported based on
an advanced renormalisation-group treatment in Ref. \cite{Finkelstein05}. On 
the other hand, numerical studies based on the diffusion Monte Carlo 
technique\cite{Senatore09} did not find any critical behaviour of 
susceptibility in a low density two-valley 2DEG.

The effects of finite thickness of the inversion layer on the quasi-2DEG 
susceptibility were addressed, {\it e.g.}, in Refs. 
\cite{DasSarma05,Senatore05,Tutuc03}. In these investigations, the appropriate 
diagrammatic summations \cite{DasSarma05} or the Monte-Carlo numerical results 
\cite{Senatore05} are generalised for the quasi-2D case by including the appropriate formfactors\cite{Ando82},
whereas for interpreting the experimental findings in Ref. \cite{Tutuc03}, 
orbital effects of the in-plane magnetic field (which can be non-negligible 
for a sufficiently thick layer) were invoked.

We note that within the general framework of delocalised-electron magnetism, 
an important benchmark is provided by Stoner-type mean-field theories. 
The underlying assumption is the presence of a local repulsive interaction 
between opposite-spin electrons. In the case of a quasi-2DEG, where an 
important role is played by long-range Coulomb correlations, such an assumption
does not at the first sight appear too realistic. Yet we suggest that it 
might be worthwhile to
study it in some detail, especially as a relative simplicity of the situation 
allows for a full mean-field treatment including the electrostatic effects.
We thus aim at generalising the classical Ref. \cite{Stern72} to treat the 
magnetic response of a silicon inversion layer, where in addition to the 
long-range Coulomb interaction (which at the mean-field level is included in 
the effective electrostatic potential), contact (on-site) repulsion between 
the carriers is also taken into account.
We note that in the case of silicon, recent density functional theory 
results\cite{Cococcioni2010} suggest a rather large value of on-site 
repulsion $U_{on-site}\approx 3 eV$. This is somewhat unexpected for a 
semiconductor 
with covalent bonds, and perhaps provides an additional motivation for the
present study. The validity of our mean-field approach will be discussed 
elsewhere.      

\section{Mean-Field Description of an Inversion Layer}
\label{sec:mf}

Following Ref. \cite{Stern72}, we consider an $n$-doped Si inversion
layer, with the value of chemical potential fixed at the top of the
valence band $E_v$ of the bulk silicon. We set $E_v$ to be our zero of
energy, and we further assume the presence of an acceptor level at
$E=E_v=0$, with the volume density of acceptors $N_A$. The bottom of
conduction band bends from its bulk value $E_c \approx 1.12$eV (attained
far away from the surface, where the transverse co-ordinate $z$ is
large and positive) to a variable value $E_{cs}<0$ at the surface
($z=0$), which is controlled by the electrostatic potential
$\phi_{gate}$ at the metallic gate, 
\begin{equation}
\phi(z=0) \equiv -\frac{1}{e}(E_{cs}-E_c) =-\frac{1}{C}Q_{gate}+\phi_{gate}\,.
\label{eq:phigate}
\end{equation}
Here, we assume that the potential $\phi(z)$ vanishes deep within the
bulk (more precisely, at $z>z_d$, where $z_d$ is the depletion layer width);
$-e$ is the electron charge, $C$ is the capacitance per unit area of
the oxide layer, and $Q_{gate}$ is the (positive) surface charge density
at the gate, which exactly compensates the induced charges in the
semiconductor: $Q_{gate}=e(n_\uparrow+n_\downarrow+N_A z_d)$, with
$n_\uparrow$ ($n_\downarrow$) being the density of the 2DEG carriers
with spin up (down). The constant difference in work functions is
absorbed into $\phi_{gate}$. The value of $\phi$ at $z>0$ is found
from the (self-consistent) Poisson's equation as 
\begin{eqnarray}
&&\phi(z)=\frac{1}{e}(E_c-E_{cs})-\nonumber \\
&&-\frac{4\pi e}{\epsilon}(N_A z_d+n_\uparrow+n_\downarrow) z +
\frac{2 \pi e}{\epsilon}z^2 N_A 
+ \nonumber \\ 
&&\!\!\!\!\!\!\!\!\!\!\!+\frac{4\pi e}{\epsilon}\int_0^z dz' \int_0^{z'} 
\left[n_\uparrow \psi^2_\uparrow(z'')+n_\downarrow \psi^2_\downarrow(z'')
\right] dz''\,,
\label{eq:phi}
\end{eqnarray}
where $\epsilon\approx 11.9 $ is the static dielectric constant of  bulk
Si. The wave-functions $\psi_\alpha$ (with $\alpha=\uparrow,
\downarrow$) of the transverse motion are the ground-state
eigenfunctions of the effective mean field one-dimensional Hamiltonian
(we recall that in the underlying quantum mechanical problem
transverse carrier motion separates from the in-plane one, the latter
being free):  
\begin{eqnarray}
&&\!\!\!\!\!\!\!\!\!\!\!\!\!\!\!\!\!\!\!\!\!\!\!\!\!{\cal H}_\alpha\!\!=\!\!E_c- \frac{\hbar^2}{2m_\parallel}\frac{\partial^2}{\partial z^2}- 
e \phi(z)+
Un_{-\alpha}\psi^2_{-\alpha}(z)-\frac{1}{2}H\sigma^z_{\alpha \alpha}\,, 
\label{eq:Ham}\\
&&{\cal H}_\alpha\psi_\alpha=E_{0\alpha}\psi_\alpha \,.
\label{eq:Schroed}
\end{eqnarray}
Here, $m_\parallel$ is the transverse effective mass [for Si-(100)
surface, $m_\parallel \approx 0.916$ in the units of free electron mass
$m_e$], $H$ is the in-plane magnetic field in units of $g \mu_B$ (bare
$g$-factor times Bohr magnetone), and $\sigma^z$ is the Pauli
matrix. Orbital effects of the
in-plane field, which have been discussed
elsewhere\cite{Tutuc03,DasSarma2000},  are 
not included in our ${\cal H}$. $U$ is the contact repulsion 
(or an $s$-wave scattering)
potential, which is related to the on-site repulsion $U_{on-site}$ on
the underlying  
discrete lattice roughly as $U=a^3 U_{on-site}$ with $a \approx 5.43$ \AA ~the lattice
period.  While for the  Si-(100) surface   two equivalent 
electron valleys should be taken into account, here we assume that the
Hamiltonian is diagonal in the valley index. In writing
Eqs. (\ref{eq:Ham}--\ref{eq:Schroed}), we assumed that
only the ground states $E_{0\alpha}$  (for $\alpha=\uparrow,\downarrow$)
can be populated by the carriers (electric quantum limit \cite{Stern72}).\footnote{Generalisation for the high-density region where several transverse-motion levels $E_n$ are active is straightforward.} Then the respective carrier densities at zero temperature are 
given by
\begin{equation}
n_\alpha =-2 \nu E_{0\alpha}  \theta (-E_{0\alpha})\,,\,\,\,\nu=\frac{m_\bot}{2\pi \hbar^2}\,,
\label{eq:nalpha}
\end{equation}
where the factor 2 accounts for the valley degeneracy, 
$\theta$ is the Heavyside function, and $m_\bot \approx 0.190 m_e$ is the effective 
mass of the in-plane motion. The net carrier density and magnetisation (in units of $\mu_B$) are then
\begin{equation}
n= n_\uparrow+ n_\downarrow\,,\,\,\,M=\frac{n_\uparrow- n_\downarrow}{2}\,.
\label{eq:nm}
\end{equation}
The self-consistent mean field scheme is completed
by an equation for the depletion layer width $z_d$, viz..
\begin{equation}
\frac{2 \pi e^2}{\epsilon}N_A z_d^2 = E_c-E_{cs}- \frac{4 \pi e^2}{\epsilon}
(n_\uparrow z_\uparrow + n_\downarrow z_\downarrow)\,,
\label{eq:zd}
\end{equation}
where $z_\alpha=\int_0^\infty z \psi^2_{\alpha}(z) dz$ is the average value of
$z$ for spin-up or spin-down carriers. 

In the $U=0$, $H=0$ case using an {\it ansatz}\cite{Stern72,Fang1966}
\begin{equation}
\psi(z)=\sqrt{\frac{b^3}{2}}z \exp(-bz/2)
\label{eq:Stern}
\end{equation}
with an appropriate $b$ yields the values of $E_0$ which agree quite well 
with numerical 
solution of Eqs. (\ref{eq:Ham}--\ref{eq:Schroed}). We find a similar 
behaviour in the $U>0$, $H=0$ unpolarised case, where the equation for $b$ now 
reads 

\begin{equation}
\frac{\hbar^2 b^3}{4 m_\parallel} +\frac{3}{64}Un b^2- \frac{12 \pi e^2}{\epsilon} \left(N_A z_d + \frac{11}{32}n \right) =0\,.
\end {equation}

The addition of the second term on the l.\ h.\ s.\ results in a
slight decrease of the value
of $b$ (and hence in an increase of $z_\uparrow=z_\downarrow = 3/b$) in
comparison to the non-interacting case, as can be expected on
general grounds.

\section{Magnetic susceptibility}
\label{sec:susce}

Conventional Stoner picture in the uniform three-dimensional case
involves the energy shifts of the two spin subbands under the combined
effect of $H$ and the Hartree field {\it without} changing the wave
functions. When using the {\it ansatz} (\ref{eq:Stern}) in the case of
an inversion layer in the electric quantum limit, this approach yields  
the following expression
for the magnetic susceptibility:
\begin{equation}
\chi_{St} \equiv \left. \frac{\partial M}{\partial H} \right|_{H\rightarrow 0}= 
\frac{\nu}{1-\frac{3}{8}Ub\nu}\,.
\label{eq:2dstoner}
\end{equation}
We note that $\chi$ depends on the carrier density $n$, due to the
presence of an $n$-dependent quantity $b$ in the denominator. This is
in turn due to the fact that $U$ is a three-dimensional contact
interaction, whereas the two-dimensional contact repulsion between the
2DEG carriers is in general obtained by integrating over $z$ as
\begin{equation}
U_{2D}= U \int_0^\infty \psi^2_\uparrow (z) \psi^2_\downarrow(z) d z\,.
\label{eq:U2D}
\end{equation}
In the unpolarised case, while also using the {\it ansatz}
(\ref{eq:Stern}), one obtains $U_{2D}=3 U b /16$. 
Since $b$ increases with $n$ \cite{Stern72}, so does the
susceptibility, Eq. (\ref{eq:2dstoner}) (see below, Fig. \ref{fig:susce}).

It is however obvious that for $U>0$ the transverse-motion wave 
functions $\psi_\alpha(z)$ do change when spin polarisation is
induced. In order to clarify this, we will first assume that for each spin 
projection, only the ground state $E_{0\alpha}$ of $z$-axis motion  is 
populated with carriers. 
While this assumption is quantitatively adequate only for smaller densities 
below $3\cdot 10^{12}$ cm$^{-2}$, it allows an easy insight into the generic 
qualitative behaviour which persists also in a complete (and more cumbersome) 
multi-level picture needed at higher densities, which we will discuss later.

\begin{figure}
\includegraphics[width=.49\textwidth]{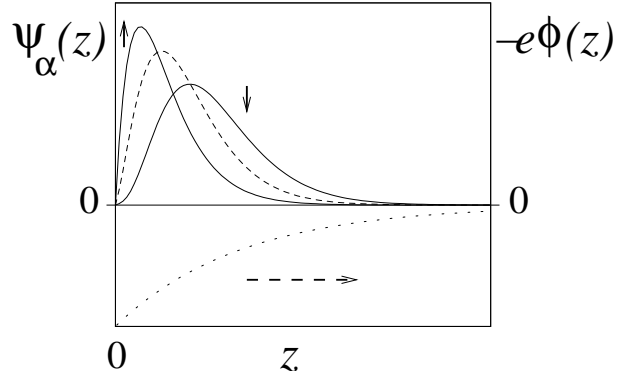}
\caption{\label{fig:scheme} Schematic representation of the effect of 
spin polarisation on the transverse-motion wave-functions at $U>0$. 
While in the unpolarised case spin-up and -down wavefunctions coincide 
(dashed line), this is no longer true at $M>0$ (solid lines). Dotted line 
represents the electrostatic potential energy, $-e\phi(z)$. }  
\end{figure}

In the presence of spin polarisation, the Hartree field of spin-up (majority)
electrons creates a potential bump for the spin-down electrons [see
Eq. (\ref{eq:Ham})], with the result that spin-down electrons are
pushed towards larger $z$. This new spin-down electron distribution
in turn creates a (smaller) bump for the spin-up electrons at these
larger values of $z$, as a result pushing the spin-up electrons closer to the
surface. This situation is shown schematically in
Fig. \ref{fig:scheme}. While for simplicity we show the electrostatic
potential $\phi(z)$ as a single dotted line (as if it does not change
when polarisation arises), in reality it is affected by the change in
wave functions [see Eq. (\ref{eq:phi})], and the resultant new
$\phi(z)$ must be fed back self-consistently into Eq. (\ref{eq:Ham}).
The interaction energy decreases with decreasing wavefunction overlap,
so that sufficiently large $U$ would imply, at the mean field level, 
a tendency towards ferromagnetism which (unlike in the usual Stoner picture) 
involves a degree of spatial separation of opposite-spin carriers along the 
transverse direction. 
 
In order to probe this picture variationally, we write instead of Eq. (\ref{eq:Stern}):
\begin{equation}
\psi_\alpha(z)=\sqrt{\frac{b^3_\alpha}{2}} a_\alpha z \sqrt{1+ \gamma_\alpha b_\alpha^2 z^2} \exp(-b_\alpha z/2)\,.
\label{eq:wf160}
\end{equation}
Here the parameters $\gamma_\alpha$ and $b_\alpha$ control the
behaviour of wave function at small and large values of $z$
respectively, whereas $a_\alpha$ is a normalisation coefficient.
In the absence of polarisation, the coefficients here do not depend on
the spin index $\alpha$. The values of $b$ and $\gamma$ can be found
by minimising the appropriate thermodynamic potential $G$. It is somewhat
simpler to work this out at a fixed total charge $n+N_Az_d$ and hence
fixed $E_{cs}$ [cf. Eq. (\ref{eq:phigate})], as opposed to
fixed $\phi_{gate}$ (while the actual situation corresponds to fixing 
$\phi_{gate}$, the two approaches are equivalent for $M=0$ or for an
infinitesimal $M$ needed when calculating susceptibility). In the unpolarised
case, we have
\begin{eqnarray}
&&\!\!\!\!\!\!\!\!\!\!\!\!\!\!\!\!\!\!\!G= -n\int_0^\infty \psi(z) \frac{\hbar^2}{2 m_\parallel}
  \frac{\partial^2}{\partial z^2} \psi(z) dz + \frac{n^2}{8\nu} + n
  E_c+\nonumber \\ 
&&\!\!\!\!\!\!\!\!\!\!\!+\frac{Un^2}{4}
  \int_0^\infty\psi^4(z) dz -  \frac{en}{2}\int_0^\infty \psi^2(z)
  \phi(z)dz-\nonumber \\
&&\!\!\!\!\!\!\!\!\!\!\!-\frac{eN_A}{2}
  \int_0^{z_d} \phi(z)dz -\frac{1}{2}(n+N_A z_d)(E_c-E_{cs}) 
\label{eq:G}
\end{eqnarray} 
Here, the first four terms are transverse and in-plane kinetic
energy of the carriers, the band-gap contribution and the interaction
energy;  
the next two terms are the electrostatic energies of carriers and
acceptors [see Eq. (\ref{eq:phi})], and the last term corresponds to
the appropriate choice of thermodynamic potential for the case of
fixed $E_{cs}$ as explained above. The values of $z_d$ and $n$ are
found self-consistently from Eqs. (\ref{eq:zd}) and
(\ref{eq:nalpha}--\ref{eq:nm}), respectively, where in
Eq. (\ref{eq:nalpha}) one has to substitute for $E_{0\alpha}$ the variational 
energy value,
\begin{eqnarray}
&&\!\!\!\!\!\!\!\!\!\!\!\!\!\!\!\!\!E_0=-\int_0^\infty \psi(z) \frac{\hbar^2}{2 m_\parallel}
  \frac{\partial^2}{\partial z^2} \psi(z) dz +E_c+ \nonumber \\
&&\!\!\!\!\!\!\!\!\!\!\!\!\!\!\!\!\!+\frac{Un}{2} \int_0^\infty\psi^4(z) dz-\frac{e}{2}\int_0^\infty
  \psi^2(z) \phi(z)dz\,.
\label{eq:E0}
\end{eqnarray}
Further details can be found in the Appendix.

We note that even in an unpolarised case at $U=0$, the wavefunction
(\ref{eq:wf160}) represents an improvement in comparison to 
Eq. (\ref{eq:Stern}), as indicated by a non-zero value of the
coefficient $\gamma>0$ found from the energy minimisation. With
increasing $U$, the value of $\gamma$ decreases but stays positive
throughout physically relevant range of parameter values. When a
polarisation is induced, $M \neq 0$, the parameters of the wave
function (\ref{eq:wf160}) depend on the spin index. As long as
polarisation is small, $M/n \ll 1$, we write to leading order

\begin{equation}
b_\alpha = (1 \pm \beta) b\,,\,\,\,\,\gamma_\alpha = \gamma \pm \delta\,.
\label{eq:betadelta}
\end{equation}

Collecting the leading-order corrections to the self-consistent  
electrostatic energies of carriers and acceptors, as well as to the
variational energies $E_{0\alpha}$ of the transverse motion, enables
one to evaluate
the susceptibility. The resultant expression is somewhat cumbersome and is given
in the Appendix [see Eq. (\ref{eq:suscevar})]. 

Results are presented in Fig.\ref{fig:susce} (a), where the dashed line
corresponds to the variational calculation based on
Eq. (\ref{eq:wf160}). The computed value of $\chi$ exceeds that of the
non-interacting case (Pauli susceptibility, $\chi \equiv \nu$ in our
notation), and increases with increasing $n$. One can see that the
variational calculation somewhat overestimates the tendency
toward ferromagnetism, suggesting a transition at $n_v \approx 1.9
\cdot 10^{13}$ cm$^{-2}$ ,
whereas the numerical solution of mean-field equations [including
the Schr\"{o}dinger equation (\ref{eq:Schroed}); see solid line in 
Fig. \ref{fig:susce} (a)] yields the
critical value $n_1 \approx 5.4 \cdot 10^{13} $ cm$^{-2}$. Importantly, 
the ``variational Stoner''
approach, using Eqs. (\ref{eq:Stern}) and (\ref{eq:2dstoner}) (dotted
line in Fig. \ref{fig:susce} (a) ), strongly underestimates $\chi$ and
misses the transition. Indeed, 
the ratio $\chi_{St}/\nu$ slowly increases with density, 
reaching the value of 2  only at $n \sim 5 \cdot 10^{14}$ cm$^-2$ (this is for 
$U=10^{-33}$
  erg/cm$^{3}$, as used in Fig. \ref{fig:susce}) ; on the other hand,
we find that for  $\chi_{St}$ to diverge at $n=n_1$, one must increase 
the value of $U$ to $U\approx 4.0 \cdot 10^{-33}$
  erg/cm$^{3}$, {\it i.e.}, by a factor of 4.

If one neglects the effect of  wavefunction change,  depicted schematically in 
Fig. \ref{fig:scheme}, the Stoner criterion of ferromagnetism is
determined precisely by the
divergence of  $\chi_{St}$, Eq. (\ref{eq:2dstoner}) as
\begin{equation}
2\nu U_{2D}  > 1\,
\end{equation}
{\it i.e.,} takes on a purely-2D form.
We see that treating the $z$-axis motion properly (which includes taking into
account interaction-induced wave functions change in the presence of spin 
polarisation)
{\it relaxes} the Stoner criterion.

\begin{figure*}
\includegraphics[width=\textwidth]{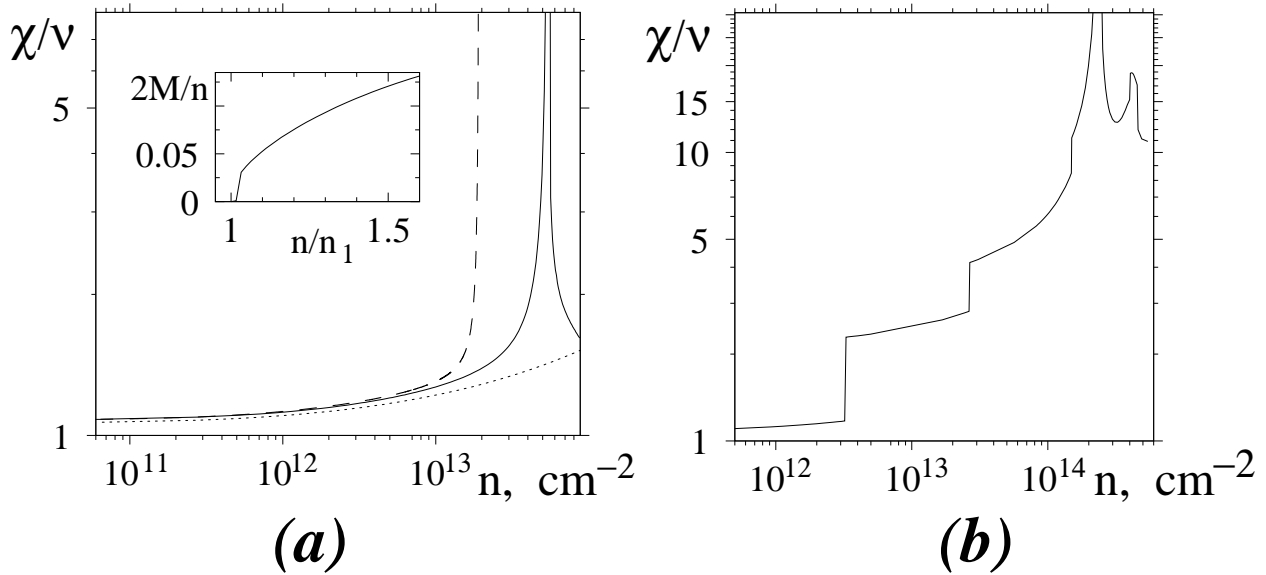}
\caption{\label{fig:susce} Magnetic susceptibility $\chi$ as a
  function of inversion-layer carrier density $n$. (a) Results assuming 
single populated level for the $z$-axis motion. Solid, dashed, and
  dotted line correspond respectively to the numerical solution of the 
mean-field  equations, variational calculation based on Eq. (\ref{eq:wf160}),
  and the ``Stoner'' value calculated using Eqs.(\ref{eq:Stern}) and
  (\ref{eq:2dstoner}). The inset shows the numerical solution result
  for the degree of spin polarisation $2M/n$, which differs from zero
  above the corresponding critical value, $n_1 \approx 5.4 \cdot
  10^{13}$ cm$^{-2}$.  The
  parameter values are $N_A=10^{15}$ cm$^{-3}$ and $U=10^{-33}$
  erg/cm$^{3}$, the latter corresponding roughly to $U_{on-site}
  = 4$eV. (b) Numerical results of the full multi-level calculation, showing
transition at  $n_0\approx 2.4 \cdot 10^{14}$
cm$^{-2}$.  } 
\end{figure*}

The inset in Fig. \ref{fig:susce} (a) shows the numerical results for the 
behaviour of spin polarisation degree $2M/n$ above the transition. 
It is seen that $M$ increases continuously, and the transition appears to be 
second-order. This is an important difference from a conventional Stoner 
theory in two dimensions where (in the case of a perfectly flat 
density of state and at a fixed $n$, and 
at the level of the present treatment) the 
total energy at the critical point does not change as $2M/n$ is varied from 
0 to 1, and a fully spin-polarised state is stabilised beyond the transition, 
resulting in a magnetisation jump. In reality, however, the opposite-spin 
transverse wave functions overlap decreases with increasing $M$ (cf. Fig. 
\ref{fig:scheme}), hence the effective two-dimensional repulsion, 
Eq. (\ref{eq:U2D}) and the Hartree field also 
decrease, resulting in a smooth transition.

As explained above, the results plotted in Fig. \ref{fig:susce} (a) are only 
indicative,
as in reality at those densities where the critical behaviour is suggested more
then one quantum-mechanical level of the $z$ axis motion is populated. In Fig.  
\ref{fig:susce} (b), we show the results of full numerical calculation of 
susceptibility. These are obtained by solving Eqs. 
(\ref{eq:phi}--\ref{eq:nalpha}), 
generalised for the case of multiple active levels. With increasing density, 
carriers begin to populate the 2nd, 3rd, 4th, and 5th level at 
$n \approx 3.2 \cdot 10^{12}$ cm$^{-2}$, $n \approx 2.6 \cdot 10^{13}$ cm$^{-2}$, 
$n \approx 1.5 \cdot 10^{14}$ cm$^{-2}$, 
and $n \approx 4.0 \cdot 10^{14}$ cm$^{-2}$, respectively. 
This gives rise to the vertical steps in susceptiblity, which are 
clealy visible in Fig.  
\ref{fig:susce} (b). Aside from these, the overall 
behaviour is qualitatively similar to the single-level results of 
Fig. \ref{fig:susce} (a), 
although the actual {\it mean-field critical density for the ferromagnetic transition,} $n_0\approx 2.4 \cdot 10^{14}$
cm$^{-2}$,  is larger than $n_1$. At $n = n_0$, 
we find 4 active levels for each spin projection. 
The increase of critical density as compared to the single-level results 
of Fig.
\ref{fig:susce} (a) is due to a broader spatial spread of higher-level 
wavefunctions (which reduces both the effective 2D interaction and the 
wavefunction change under the effect of magnetic 
field). Note that in Fig. \ref{fig:susce},  variational 
results are presented for the single-level case only, as the required 
calculations become more cumbersome in the case of multiple occupied levels.

\subsection{Phenomenological treatment of correlations near MIT}
\label{subsec:dolgo}

So far, we found that the magnetic susceptibility $\chi$ increases with 
increasing density $n$, with a transition to ferromagnetism deep
inside the high-density metallic region. An important underlying assumption was
that the in-plane carrier 
motion is free, which is no longer adequate for smaller densities close to MIT.
Indeed, in this region the dominant role is played by the in-plane
Coulomb correlations, as indicated by the measure of the interaction
strength, $r_s=m_\bot e^2/(\epsilon \hbar^2 \sqrt{\pi n})$, exceeding unity.
Very recently, it has been suggested\cite{Dolgopolov15} that the
available data at low densities can be described by a phenomenological
model 
of a 2D Fermi gas with a renormalised in-plane mass:
\begin{equation}
\tilde{m}_\bot = m_\bot \frac{n}{n-n_c}\,.
\label{eq:dolgo}
\end{equation}
The Pauli-like in-plane magnetic susceptibility would then be\cite{Dolgopolov15}
\begin{equation}
\chi_{P}=\tilde{\nu}\,,\,\,\,\,\tilde{\nu}=\frac{\tilde{m}_\bot}{2\pi \hbar^2}\,.
\label{eq:Pauli}
\end{equation}
When $n$ approaches $n_c$ from above, $\chi_P$ increases because of the 
effective band narrowing, which does not necessarily imply a magnetic 
instability (in agreement also with Ref. \cite{Finkelstein05}).

\begin{figure}
\includegraphics[width=.49\textwidth]{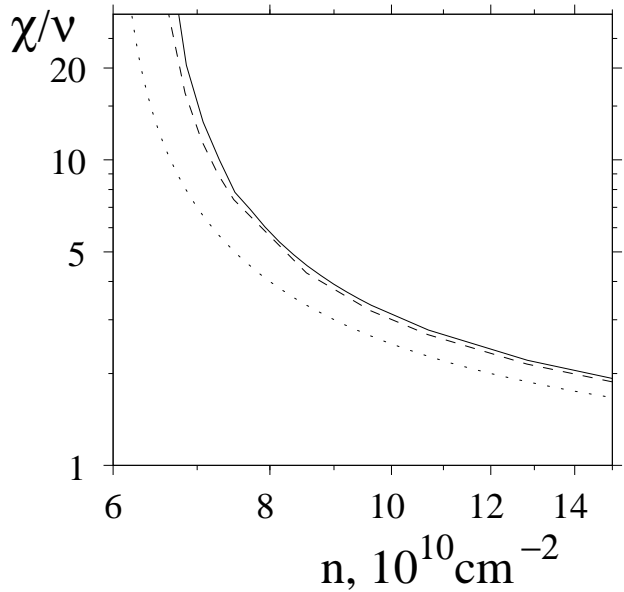}
\caption{\label{fig:dolgo} Magnetic susceptibility $\chi$ divided by
  the bare density of states $\nu$ [see Eq. (\ref{eq:nalpha})]  at low
  densities. In-plane carrier mass is renormalised according to 
  Eq.(\ref{eq:dolgo}). Solid, dashed, and
  dotted line correspond respectively to the numerical solution of the 
mean-field  equations,  ``Stoner'' susceptibility calculated using 
Eqs.(\ref{eq:Stern}),(\ref{eq:2dstoner}) with $\nu \rightarrow \tilde{\nu}$, and
Pauli susceptibility  (\ref{eq:Pauli}). The
  parameter values are $N_A=10^{15}$ cm$^{-3}$, $U=10^{-33}$
  erg/cm$^{3}$, and $n_c=6 \cdot 10^{10}$ cm$^{-2}$.} 
\end{figure}

Since there appears to be no reason to expect that the {\it transverse} 
motion of carriers is affected by the correlations, we suggest that the 
effects of a short-range $U$ can be taken into account by substituting 
$\tilde{\nu} \rightarrow \nu$ in our treatment as described above. It is readily
seen that, for example, the variational Stoner susceptibility 
(\ref{eq:2dstoner}) would
diverge at $n=n_S$, corresponding to a zero of the denominator. For
the purpose of an estimate we may use a non-interacting ($U=0$) value
of $b(n)$ \cite{Stern72},  
yielding an equation
\begin{equation}
\frac{3}{4}U \nu \frac{n_S}{n_S-n_c}
\left[ \frac{6 m_\parallel \pi e^2}{\epsilon \hbar^2}\left( N_A
  z_d+\frac{11}{32} n_S \right) \right]^{1/3}=1\,. 
\end{equation}
This suggests that a ferromagnetic state is stabilised at $n_c<n<n_S$.
Further refinement can be obtained by using a more adequate variational 
{\it ansatz} such as (\ref{eq:wf160}), or by solving the mean-field 
equations numerically. This is illustrated in Fig. \ref{fig:dolgo}.

Indeed, we see that while $\chi_P$, Eq. (\ref{eq:Pauli}), does not show
any divergence above $n_c$, both the variational Stoner result and the numerical
solution suggest a transition (at $n_* \approx 1.08 n_c$ for numerical and at 
$n_S\approx 1.07 n_c$ for variational Stoner; this is for the value of $U$ used 
in Fig. 
\ref{fig:dolgo}). While the two follow
each other rather closely in this regime, the numerical result for
$\chi$ (and hence for the density value at transition) is larger than
the variational one, mostly due to a pronounced difference between the 
exact transverse wavefunction and the variational one 
[given by Eq. (\ref{eq:Stern})] in this regime. As
explained above, the transition is expected to be second-order. Indeed, 
numerical results imply that $M(n)$ is continuous at the transition point, 
$n=n_*$,
and rapidly increases with decreasing density,
reaching full spin polarisation at $n_F>n_c$. Full spin polarisation is then
sustained in the region $n_c<n<n_F$.

Alternatively, taking into account the accuracy of experimental data, it 
appears possible that the actual MIT takes place at the point $n_F$, whereas
$n_c$, which enters Eq. (\ref{eq:dolgo}), plays the role of an 
asymptotically defined parameter (a point where the effective mass would 
have diverged).   

\section{Discussion}

We constructed a Stoner-type theory of magnetism for electrons in an 
inversion layer, addressing both the behaviour of the system in the metallic 
high-density region and
the correlated low-density regime immediately above the metal-insulator transition.  While we specifically aimed at describing 
Si-(100) MOSFETs, our results are expected to be qualitatively
relevant for other 2D electron systems of finite thickness. These general
conclusions are: (i) At higher densities, proper treatment requires taking 
into account the wave function change under the applied in-plane magnetic field 
(see Fig. \ref{fig:scheme}). This
effect leads to an increased susceptibility in the paramagnetic state, and 
enhances
the tendency toward ferromagnetism. As a result, the Stoner criterion is 
somewhat relaxed in comparison to a more conventional treatment, which in turn
may give rise to a ferromagnetic transition. (ii) Even though the overall 
behaviour  at low densities is dominated by long-range Coulomb forces, 
the magnetic susceptibility 
can be significantly affected by the on-site carrier repulsion, which is of 
course always present.  

We close with a brief summary of our results for a silicon MOSFET.
In the high-density metallic regime, when one 
neglects the in-plane Coulomb correlations 
(main body of Sec. \ref{sec:susce}), the relaxation of the Stoner criterion
translates into a lower 
value of critical density corresponding to a ferromagnetic transition (in 
comparison to the conventional Stoner approach). 
Still, the transition to ferromagnetism occurs 
at rather large $n$, and it is far from clear whether the corresponding 
values can be reached in case of Si-MOSFETs. However, the increase of 
susceptibility with increasing $n$ should perhaps be observable at large yet
readily accessible 
densities, $n \gtrsim 10^{12}$ cm$^{-2}$.

Describing the magnetic response of the system at lower densities close 
to the metal-insulator transition requires taking into account the 
strongly-correlated nature of the 2DEG in this limit. This was carried out 
at a phenomenological level in sec. \ref{subsec:dolgo}, and results suggest
a ferromagnetic instability above the metal-insulator transition. This finding
contributes to the on-going debate in the literature\cite{Reznikov03,Kravchenko06}.

The mean-field transition to ferromagnetism is found to be continuous in all 
cases, and a sublinear behaviour of $M(H)$ at larger $H$ is anticipated 
(due to decreasing wavefunction overlap).
Another important and novel feature of our approach 
is the spatial separation (in the transverse direction)
which arises between opposite-spin carriers in the ferromagnetic case or 
whenever spin polarisation is induced (the average values of the transverse
coordinate of spin-up and -down electrons then 
differ, $z_\uparrow < z_\downarrow$). This issue will be addressed in more 
detail elsewhere.      

\section{Acknowledgements}
The author takes pleasure in thanking R. Berkovits, K. A. Kikoin, 
B. D. Laikhtman, S. V. Kravchenko, I. Shlimak, and L. D. Shvartsman 
for enlightening discussions. This work was supported by the Israeli 
Absorption Ministry.

\appendix

\section{Details of the variational calculation based on Eq. (\ref{eq:wf160})}

As a first step, one has to determine the values of the wavefunction 
parameters $b$ and $\gamma$ in the unpolarised case by minimising the 
thermodynamic potential $G$, Eq. (\ref{eq:G}). Evaluating all terms in
Eqs. (\ref{eq:G}) and (\ref{eq:E0}) in the usual limit of 
$z_\alpha \ll z_d$ and differentiating, we find 
\begin{eqnarray}
&&  \!\!\!\!\!\!\!\!\!\!\!\!\!\!\!\!\!\!\!\! b^2 \frac{\hbar^2 a^2}
{2 m_\parallel} \left[\Gamma-\frac{1}{2}+ 2\gamma\right]  + 
\frac{3}{2^{12}}bnU (69
\xi^2-210 \xi + 205) - \nonumber \\ 
&&  \!\!\!\!\!\!\!\!\!\!\!\!\!\!\!\!\!\!\!\!- 
\frac{12 \pi e^2 N_A z_d \xi}{\epsilon b} -\frac{3 \pi
  e^2 n}{512 b \epsilon}(435 \xi^2\!\!-286 \xi+555)\!\!=\!\!0\,,
\label{eq:b160} \\
&&  \!\!\!\!\!\!\!\!\!\!\!\!\!\!\!\!\!\!\!\! b^2\frac{\hbar^2 a^2}
{4 m_\parallel} \left[-2-6a^2+24 \gamma a^2+
  \frac{\Gamma}{2\gamma}(1+24 \gamma a^2) + \frac{\Omega}{2 \gamma^2}
  \right]= \nonumber \\
&&  \!\!\!\!\!\!\!\!\!\!\!\!\!\!\!\!\!\!\!\! =\frac{3}{256} bnU a^4 
(69 \xi -105) + \frac{96 \pi e^2 N_A z_d}{b
  \epsilon}a^4 + \nonumber \\
&&  \!\!\!\!\!\!\!\!\!\!\!\!\!\!\!\!\!\!\!\! +\frac{3}{32} 
\frac{\pi e^2 n}{b \epsilon}a^4 (435 \xi-143)\,.
\label{eq:gamma160}
\end{eqnarray}
Here, the quantities $\Gamma$ and $\Omega$ depend only on $\gamma$, and are
expressed in terms of sine and cosine integrals:
\begin{eqnarray}
\Gamma = \frac{1}{\sqrt{\gamma}} \left({\rm ci}\frac{1}{\sqrt{\gamma}} \sin
\frac{1}{\sqrt{\gamma}}-{\rm si}\frac{1}{\sqrt{\gamma}} \cos
\frac{1}{\sqrt{\gamma}}\right) \,,\\
\Omega={\rm ci}\frac{1}{\sqrt{\gamma}} \cos
\frac{1}{\sqrt{\gamma}}+{\rm si}\frac{1}{\sqrt{\gamma}} \sin
\frac{1}{\sqrt{\gamma}}\,;  
\end{eqnarray}
$a=(1+12 \gamma)^{-1/2}$ is the normalisation coefficient in
Eq. (\ref{eq:wf160}) at $m=0$, and $\xi$ is defined according to
$\xi=a^2(1+20 \gamma)$, so that $z_\uparrow=z_\downarrow=3 \xi/b$.
Eqs. (\ref{eq:b160}--\ref{eq:gamma160}) must be solved numerically,
together with Eqs. (\ref{eq:zd}) and (\ref{eq:nalpha}--\ref{eq:nm}).

When an infinitesimal in-plane magnetic field $H$ is applied, the
spin-up and -down wavefunctions begin to differ from each other
according to Eq. (\ref{eq:betadelta}). This effect, along with 
the Zeeman splitting of the spin subbands, contributes to the magnetisation,
\begin{eqnarray}
&&  \!\!\!\!\!\!\!\!\!\!\!\!\!\!\!\!\!\!\!\! M=
  \left[
(69 \xi^2-210 \xi+205)\frac{3bnU\nu}{2048} \beta +(69 \xi -105) a^4 
\times \right.\nonumber \\
&&  \!\!\!\!\!\!\!\!\!\!\!\!\!\!\!\!\!\!\!\!\!\!
   \left.\times\frac{3 bnU\nu}{128} \delta 
\!+\!\nu H\right] \left[1\!-\!
    \frac{3bU\nu}{512} (69 \xi^2\!\!-\!210 \xi \!+\!205) \right]^{-1}\!.
\label{eq:mvar}
\end{eqnarray}

The thermodynamic potential acquires a correction, $G \rightarrow G+
\Delta G$, which is quadratic in $\beta$, $\delta$, and $H$. The
values of $\beta$ and $\delta$ can be found by minimising $\Delta G$,
yielding magnetisation $M$ [with the help of Eq. (\ref{eq:mvar})], and
hence the susceptibility $\chi$. 
A straightforward, if somewhat tedious, calculation yields
\begin{equation}
\chi=\left[1-\frac{3bU\nu}{512}(69 \xi^2-210\xi+205) -K\right]^{-1}
\label{eq:suscevar}
\end{equation}
Here the first two terms correspond to the Stoner susceptibility,
computed using the wavefunctions (\ref{eq:wf160}) of the unpolarised
case [instead of (\ref{eq:Stern}); cf. Eq. (\ref{eq:2dstoner})], whereas
the quantity
\begin{eqnarray}
&&  \!\!\!\!\!\!\!\!\!\!\!\!\!\!\!\!\!\!\!\! 
K=\frac{9\nu b^2 n U^2}{2^{16}(AB-C^2)} \left[\frac{A}{64}(69 \xi^2-210
  \xi+205)^2+4a^8 B\times  \right.\nonumber \\
&&  \!\!\!\!\!\!\!\!\!\!\!\!\!\!\!\!\!\!\!\!\!\!
   \left. \times(69 \xi-105)^2-\frac{a^4C}{2}(69
  \xi^2\!-\!210\xi+205)(69\xi-105) \right] \nonumber 
\end{eqnarray}
accounts for the field-induced change of the spin-up and down
wavefunctions, as illustrated by Fig. \ref{fig:scheme}. The
coefficients $A$, $B$, and $C$ are given by
\begin{eqnarray}
&&  \!\!\!\!\!\!\!\!\!\!\!\!\!\!\!\!\!\!\!\!
A= \frac{\hbar^2 a^2 n b^2}{2m_\parallel} \left[\Gamma \left(\frac{6a^2}{\gamma}
+\frac{3}{8 \gamma^2} -\frac{1}{8 \gamma^3}+ 144 a^4 \right) + \Omega \times
\right. \nonumber \\
&&  \!\!\!\!\!\!\!\!\!\!\!\!\!\!\!\!\!\!\!\!
\left. \times \left(
\frac{6a^2}{\gamma^2}+\frac{5}{8 \gamma^3}\right)-24 a^2 +\frac{1}{8 \gamma^3}-72 a^4 + 288 a^4 \gamma \right] - \nonumber \\
&&  \!\!\!\!\!\!\!\!\!\!\!\!\!\!\!\!\!\!\!\!
-\frac{3 \pi e^2 n^2}{4 \epsilon b} 
(1305 \xi -429) a^6 - \frac{2^8 \cdot 9 \pi e^2 N_A z_d n a^6}{\epsilon b} + 
\nonumber \\
&&  \!\!\!\!\!\!\!\!\!\!\!\!\!\!\!\!\!\!\!\! 
+\frac{9}{64} U b n^2 (95-69 \xi) a^6 \,,\nonumber \\
&&  \!\!\!\!\!\!\!\!\!\!\!\!\!\!\!\!\!\!\!\! 
B=  \frac{\hbar^2 a^2 n b^2}{2m_\parallel} (\Gamma-\frac{1}{2}+ 2 \gamma)-
\frac{3 \pi e^2 n^2}{2^9 \epsilon b}(975 \xi^2-2638 \xi +
\nonumber \\
&&  \!\!\!\!\!\!\!\!\!\!\!\!\!\!\!\!\!\!\!\!
+1215) + \frac{24 \pi e^2 N_A z_d n \xi}{\epsilon b} + \frac{9}{2^{11}}Ubn^2(-199 \xi^2+510 \xi - \nonumber \\
&&  \!\!\!\!\!\!\!\!\!\!\!\!\!\!\!\!\!\!\!\! -375)\,,
\nonumber \\
&&  \!\!\!\!\!\!\!\!\!\!\!\!\!\!\!\!\!\!\!\! 
C=\frac{\hbar^2 a^2 n b^2}{2m_\parallel} \left[-\Gamma \left(\frac{1}{2\gamma}+12 a^2 \right)+ 2+ 6 a^2 - 24 a^2 \gamma^2- \right.
\nonumber \\
&&  \!\!\!\!\!\!\!\!\!\!\!\!\!\!\!\!\!\!\!\! 
\left.-\frac{\Omega}{2 \gamma^2} \right] - \frac{3 \pi e^2 n^2}{64  \epsilon b} (435 \xi -751) a^4 - \frac{96 \pi e^2 N_A z_d n a^4}{\epsilon b}+
 \nonumber \\
&&  \!\!\!\!\!\!\!\!\!\!\!\!\!\!\!\!\!\!\!\! 
+\frac{45}{32}Ubn^2 a^4 \,.
\nonumber
\end{eqnarray}

\begin{thebibliography}{00}



\bibitem{Ando82} T. Ando, A. B. Fowler, and F. Stern, Rev. Mod. Phys. {\bf 54},
437 (1982).

\bibitem{Kravchenko04} S. V. Kravchenko and M. P. Sarachik, Rep. Prog. Phys.
 {\bf 67}, 1 (2004), and references therein.

\bibitem{Reznikov03} O. Prus, Y. Yaish, M. Reznikov, U. Sivan, and V. Pudalov, Phys. Rev. {\bf B67}, 205407 (2003).

\bibitem{Kravchenko06} S. V. Kravchenko, A. A. Shashkin, S. Anissimova, 
A. Venkatesan, M. R. Sakr, V. T. Dolgopolov, and T. M. Klapwijk, Ann. Phys. {\bf 321}, 1588 (2006); A. A. Shashkin, S. Anissimova, M. R. Sakr, S. V. Kravchenko, V. T. Dolgopolov, and T. M. Klapwijk, Phys. Rev. Lett. {\bf 96}, 036403 (2006).






\bibitem{Tanatar89} B. Tanatar and D. M. Ceperley, Phys. Rev. {\bf B39}, 5005 (1989).
\bibitem{Attaccalite02} C. Attaccalite, S. Moroni, P. Gori-Giorgi, and G. B. Bachelet, Phys. Rev. Lett. {\bf 88}, 256601 (2002).




\bibitem{Finkelstein05} A. Punnoose and A. M. Finkelstein, Science {\bf 310}, 289 (2005).

\bibitem{Senatore09} M. Marchi, S. De Palo, S. Moroni, and G. Senatore, Phys. Rev. {\bf B80}, 035103 (2009).






\bibitem{DasSarma05} Y. Zhang and S. Das Sarma, Phys. Rev. {\bf B72}, 075308 (2005).

\bibitem{Senatore05} S. De Palo, M. Botti, S. Noroni, and G. Senatore, Phys. Rev. Lett. {\bf 94}, 226405 (2005).

\bibitem{Tutuc03} E. Tutuc, S. Melinte, E. P. De Poortere, M. Shayegan, and R. Winkler, Phys. Rev. {\bf B67}, 241309 (2003).



\bibitem{Cococcioni2010} V. L. Campo, Jr., and M. Cococcioni, J. Phys.: Condens. Matter, {\bf 22}, 055602 (2010).

\bibitem{Stern72} F. Stern, Phys. Rev. {\bf B5}, 4891 (1972).

\bibitem{DasSarma2000} S. Das Sarma and E. H. Hwang,
  Phys. Rev. Lett. {\bf 84}, 5596 (2000).

\bibitem{Fang1966} F. F. Fang and W. E. Howard, Phys. Rev. Lett. {\bf 16}, 797 (1966).

\bibitem{Dolgopolov15} V. T. Dolgopolov, JETP Lett. {\bf 101}, 282
  (2015) [Zh. Eksp. Teor. Fiz. Pis'ma Red., {\bf 101}, 300 (2015)].

\end{thebibliography}


\end{document}